\documentclass[epj,nopacs]{svjour}
\usepackage{graphics}
\usepackage{graphicx,epsfig}
\begin{document}

\title{Fermions on deformed thick branes}

\author{W. T. Cruz\inst{1} \and A. R. Gomes\inst{2} \and C. A. S. Almeida\inst{3}}


\institute{
\inst{1}Instituto Federal de Educa\c{c}\~{a}o, Ci\^{e}ncia e Tecnologia do Cear\'{a} (IFCE), Campus Juazeiro do Norte - 63040-000 Juazeiro do Norte-Cear\'{a}-Brazil\\
\inst{2} Instituto Federal do Maranh\~ao, Campus Monte Castelo, S\~ao Lu\'is - Maranh\~ao - Brazil \\
\inst{3} Departamento de F\'{i}sica - Universidade Federal do Cear\'{a} - C.P. 6030, 60455-760 Fortaleza - Cear\'{a}-Brazil}


\abstract{
In  this work we investigate the issue of fermion localization and resonances in $(4,1)$-deformed branes constructed with one scalar field coupled with gravity.
Such models provide us branes with internal structures that turns the gravitational interaction more effective for fermions aside the brane, increasing their lifetime.
The coupling between the scalar field and spinors is a necessary condition for fermions to be localized on such branes. After performing a chiral decomposition of the five-dimensional spinor we found resonances with both chiralities. The correspondence between the spectra for left and right chirality is guaranteed and Dirac fermions are realized on the brane.
     }
\maketitle

\section{Introduction}
Brane structures were initially introduced as domain walls embedded in extra dimensions \cite{cs}. In a sense thick branes \cite{de,fase,bc} are a natural construction since the solutions can be dynamically found. More recently, interesting models have been proposed for such dynamical branes with rich structures, constructed with one \cite{Bazeia:2003aw} or more \cite{brane,es,dutra} scalar fields (see also the review given by Ref. \cite{dzhun}). The issue of localization of several fields and resonances in such branes is an interesting subject, as their investigation can guide us to which kind of brane structure is more acceptable phenomenologically. In this way, models which extend the Randall-Sundrum type II scenario \cite{rs2} were constructed and considered under the aspect of gravity localization \cite{man,mel,bgl,bh,bhr}. Here, some branes constructed with scalar fields minimally coupled with gravity are successfully considered for trapping of spin 2 massive gravitons. This means to be able  to trap zero-mode (the usual graviton), responsible for reproducing the usual Newton law of gravity on the brane in a non-relativistic limit, and massive gravity modes with finite lifetime (resonances).

Reviewing works about fermions in extra dimension models, we observed that in type I Randall-Sundrum (RS) scenario [16], spin $1/2$ and spin $3/2$ fermions are located only in the negative tension brane. However, they do not satisfy the localization conditions in the type II RS model. In these two cited models, the localization of zero mode chiral fermions is only obtained due to the introduction of a generalized Yukawa coupling. This is analogous to what happens to domain walls in the absence of gravity \cite{jackiw}. In the brane case the coupling is introduced by an interaction of fermions with a bulk scalar field. For instance, the scalar field solution can be kink like. Thus, the results with the RS model indicate that the thick brane scenarios as the most suitable for locating fermions. Thick brane models can be obtained from scalar fields with kink-like solutions. Therefore, the scalar field that couples with fermions in the Yukawa coupling will be the scalar field which the brane is made of.

Similarly, for spin 0 particles, it was shown in \cite{Abdyrakhmanov:2005fs} that massive scalar modes are not trapped in a purely gravitational interaction. To be successfully trapped on the brane, one needs an additional, non-gravitational  coupling with the scalar field that supports the brane.

Quite recently, Liu et al. \cite{Liu:2007ku} have analyzed the issue of fermion localization on a pure geometrical thick brane. Since then, the same group of authors have published a series of papers on fermion localization for several types of thick branes \cite{Liu2,Liu:2008wd,Liu4,Liu5}. Also, some of the authors of the present work considered fermionic resonances in branes with internal structure \cite{ca}. As is well known, the kind of structure of the considered brane is very
important and will produce implications concerning the methods of
field localization. In the seminal works of Bazeia and collaborators \cite{deformed,Bazeia:2003qt} a class
of topological defect solutions was constructed starting from a
specific deformation \cite{deformed} of the $\phi^4$ potential. These new solutions
may be used to mimic new brane-worlds containing internal structures
\cite{Bazeia:2003aw}. Such internal structures have implications in the
density of matter-energy along the extra dimensions \cite{brane} and
this produces a space-time background whose curvature has a
splitting. Some characteristics of such model were considered in phase transitions
in warped geometries \cite{fase}.
Also, such class of models were already studied with respect to gravity localization \cite{Bazeia:2003aw}, where it was found that zero-modes for KK gravitons exist, the solutions are stable and tachyons are forbidden.

The main goal here is to study the behavior of fermions in branes generated by the so-called two-kink solutions obtained after a deformation of a $\phi^4$ potential \cite{deformed}.
As we will see, the deformations will be very important for localization and normalization of fermionic fields. Using a well known resonance detecting method \cite{Liu4,ca,wilami1,wilami_e_ca1,wilami_e_ca2}, we analyze the massive modes arising from the dimensional reduction functions. 
We must remark that our analysis of deformed models includes the thick brane scenario, which can be recovered by changing the parameter of deformation. It is worthwhile to mention that even in the simple thick brane background no previous work considered the presence of resonances in the massive spectrum of fermions.

This paper is divided as follows. In the next section we review the deformed brane model with one scalar field.
Each particular brane solution is considered as a fixed background that is not perturbed significantly by the presence of fermions. With this approximation, in Sec. III we perform a chiral decomposition and study the presence of zero modes with left chirality. Fermionic resonances and the realization of Dirac fermions on the brane are considered in Sec. IV. The main conclusions are presented in Sec. V.

\section{Brane setup}

We start with \textbf{an} action describing one scalar field minimally coupled with gravity in five dimensions
\begin{equation}
S=\int d^{5}x \sqrt{-G}[2M^{3}R-\frac{1}{2}(\partial\phi)^{2}-V(\phi)],
\end{equation}
where $M$ is the Planck mass in $D=5$ dimensions and $R$ is the  scalar curvature.

As an Ansatz for the metric we
consider $ds^2 =e^{2A(y)}\eta_{\mu\nu}dx^{\mu}dx^{\nu} +
dy^2$, which is an extension for the Randall-Sundrum metric.
Here the bulk spacetime is asymptotically $AdS_5$, and we
have a Minkowski $M_4$ brane.

The tensor $\eta_{\mu\nu}$ is the Minkowski metric and the indices
$\mu$ and $\nu$ vary from 0 to 3. For this background we find the following equation of motion:

\begin{equation}\label{mov1}
\frac{1}{2}(\phi^{\prime})^{2}-V(\phi)=24M^{3}(A^{\prime})^{2},
\end{equation}

Here prime means derivative with respect to the extra dimension. One can find general requirements for the potential $V(\phi)$ to describe a brane world with a well defined thin limit and able to confine ordinary matter \cite{Bronnikov:2003gg}. In the presence of gravity, defining the potential as $V_p(\phi)=\frac{1}{2}\left(\frac{dW}{d\phi}\right)^2-\frac{8M^3}{3}W^2,$ and choosing the superpotential \cite{Bazeia:2003qt}
\begin{equation}\label{sup}
W_p(\phi)=\frac{p}{2p-1}\phi^{\frac{2p-1}{p}}-\frac{p}{2p+1}\phi^{\frac{2p+1}{p}},
\end{equation}
we can find the solution for $\phi_p(y)=\tanh^p(\frac{y}{p})$ and $A_p(y)$  as \cite{Bazeia:2003aw},
\newpage
\begin{small}
\begin{eqnarray}\label{a}
A_p(y)=-\frac{1}{3}\frac{p}{2p+1}\tanh^{2p}\left(\frac{y}{p}\right)- \frac{2}{3}\left(\frac{p^2}{2p-1}-\frac{p^2}{2p+1}\right)\\\nonumber
\times\biggl{\{}\ln\biggl[\cosh\left(\frac{y}{p}\right)\biggr]- \sum_{n=1}^{p-1}\frac1{2n}\tanh^{2n}\left(\frac{y}{p}\right)\biggr{\}}.
\end{eqnarray}
\end{small}
The parameter $p$ is an odd integer. The chosen form for $W_p$ allows us to obtain well-defined models when $p=1,3,5,...$, where for $p=1$ we get the standard \textbf{$\phi=\tanh(y)$} kink solution \cite{Bazeia:2003qt}.
Note that the exponential warp factor constructed with this function is localized around the membrane and that for large $y$ it approximates to the Randall-Sundrum solution \cite{Bazeia:2003aw}. The spacetime now has no singularities as we get a smooth warp factor (because of this, the model is more realistic) \cite{kehagias}.

\section{Fermionic zero-mode}

Now we consider the action for a fermion coupled with gravity, in the background given by the brane solution $\phi_p(y)$ from the
previous section.
\begin{equation}
\label{Sferm}
S=\int dx^5\sqrt{g}[\bar{\Psi}\Gamma^{M} D_{M} \Psi - f \bar\Psi\phi_p\Psi],
\end{equation}
where $f$ is the 5-dimensional Yukawa coupling.

We change the variable $y$ to $z$ following the equation
\begin{equation}
\frac{dz}{dy}=e^{-A_p}.
\end{equation}

In the new set of variables the metric is conformally flat, and the gamma matrices can be rewritten as $\Gamma^{\mu}=e^{-A_p}\gamma^{\mu}\;,\;\Gamma^5=e^{-A_p}\gamma^5$, where $\gamma^\mu$ and $\gamma^5$ are
4-dimensional flat gamma matrices in the Dirac
representation. The covariant derivatives are
\begin{equation}
D_{\mu}=\partial_{\mu}+\frac{\partial_z A_p}{2}\gamma_{\mu}\gamma^5\;,\;\;\;\;D_5=\partial_5.
\end{equation}
The former expressions allow us to write the equation of motion as
\begin{equation}
[\gamma^\mu\partial_\mu+\gamma^5(\partial_z+2\partial_zA_p)+f\phi_p e^{A_p}]\Psi(x,z)=0.
\end{equation}
Performing a chiral decomposition of  the spinor as
\begin{equation}
\Psi(x,z)=\sum_n[\psi_{Ln}(x)\alpha_{Ln}+\psi_{Rn}(x)\alpha_{Rn}(z)],
\end{equation}
the massive modes from the spinor $\Psi$ living on the brane must connect both chiralities, satisfying the equations
\begin{equation}
\gamma^{\mu}\partial_{\mu}\psi_{Ln}(x)=m\psi_{Rn}(x)\,,\,\,\gamma^{\mu}\partial_{\mu}\psi_{Rn}(x)=m\psi_{Ln}(x).
\end{equation}
From the relations
\begin{equation}
\gamma^5\psi_{Ln}(x)=-\psi_{Ln}(x), \,\, \gamma^5\psi_{Rn}(x)=\psi_{Rn}(x),
\end{equation}
we find two coupled equations for  $\alpha_{Ln}(z)$ and $\alpha_{Rn}(z)$ \cite{ca,Liu:2007ku,alejandra}:
\begin{equation}
\label{eom_psi}
[\partial_z+2\partial_zA_p + f\phi_pe^{A_p}]\alpha_{Ln} (z)= m\alpha_{Rn}(z),
\end{equation}
\begin{equation}
\label{eom_psi2}
[\partial_z+2\partial_zA_p - f\phi_pe^{A_p}]\alpha_{Rn} (z)=- m\alpha_{Ln}(z),
\end{equation}
where $m$ is the mass.

Now we investigate the possibility of localized fermionic zero massive modes.
For $m=0$, Eq. (\ref{eom_psi}) reduces to
\begin{equation}
2A'_p \alpha_{Ln}(z)+\alpha_{Ln}'(z)+fe^{A_p}\phi_p\alpha_{Ln}(z)=0.
\end{equation}
If we turn back to the variable $y$ we can write
\begin{equation}
2A'_p \alpha_{Ln}(y)+\alpha_{Ln}'(y)+f\phi_p\alpha_{Ln}(y)=0,
\end{equation}
with solution
\begin{equation}\label{sol}
\alpha_{Ln}(y)=e^{-\int_0^y dy' [f\phi_p+2 A_p(y')]}.
\end{equation}
In this solution we clearly note the contribution of the internal structure from the membrane. Remembering that
$\phi_p$ and $A_p$ depend on odd integer numbers, we will see that the value of $p$ will be determinant in order to obtain a finite solution.
With the explicit expressions for $A(y)$ and $\phi_p(y)$ in Eq. (\ref{sol}), and proceeding as Ref. \cite{kehagias}, we were able to find the relation $f>\frac{8p}{-3+12p^2}$ between the coupling constant $f$ and the parameter $p$ for $\alpha_{Ln}(y)$ to be finite.
It is important to point out that the relation between $f$ and $p$ results from the coupling of the fermion $\Psi$ with the particular kind of membrane solution, introduced in the action as $f\overline{\Psi}\phi_p\Psi$. From this relation we can see how the brane deformations influence the zero mode localization.
For $m=0$, Eq. (\ref{eom_psi2}) has the solution $\alpha_{Rn}(y)=e^{\int dy' [f\phi_p(y')-2 A_p(y')]}$. From the solutions for $A(y)$ and $\phi_p(y)$ we conclude that there is no localized right chiral zero-mode.

\section{Fermionic Resonances and Dirac fermions}
Now to complete our investigation of the presence of spinorial fields in the 4-dimensional membrane, we consider the Dirac massive equation. Since we are interested in massive localized states, we transform the equation of motion for fermions in a Schr\"odinger-like equation. For fermions in a domain wall there is a similar analysis \cite{peter}. Double walls were studied in  \cite{alejandra}.
With the transformations
$\alpha_{Ln}(z)=\overline{\alpha}_{Ln}(z) e^{-2A_p}$ and $\alpha_{Rn}(z)=\overline{\alpha}_{Rn}(z) e^{-2A_p}$, Eqs. (\ref{eom_psi}) and (\ref{eom_psi2}) result in
\begin{equation}\label{schro}
[-\partial^2_z +V_p^L]\overline{\alpha}_{Ln}(z)=m^2\overline{\alpha}_{Ln}(z),
\end{equation}
\begin{equation}\label{schro2}
[-\partial^2_z +V_p^R]\overline{\alpha}_{Rn}(z)=m^2\overline{\alpha}_{Rn}(z),
\end{equation}
where the Schr\"odinger potentials are
\begin{equation}
V_p^L=-f\partial_z\phi_p e^{A_p}- f\phi_p e^{A_p}\partial_z A_p+f^2\phi_p^2 e^{2A_p}
\end{equation}
\begin{equation}
V_p^R=+f\partial_z\phi_p e^{A_p}+ f\phi_p e^{A_p}\partial_z A_p+f^2\phi_p^2 e^{2A_p}.
\end{equation}
Due to the change \textbf{of} variables from $y$ to $z$ we have no explicit form for the potentials $V_p^L$ and $V_p^R$. From the numerically known potential we can use the Numerov numeric method \textbf{\cite{numerov}} to solve the Schr\"odinger equations. A careful comparison of this method to some other approaches is presented in Ref. \cite{bgl}.
\begin{figure}\centering
\includegraphics[width=4.2cm,height=2.5cm]{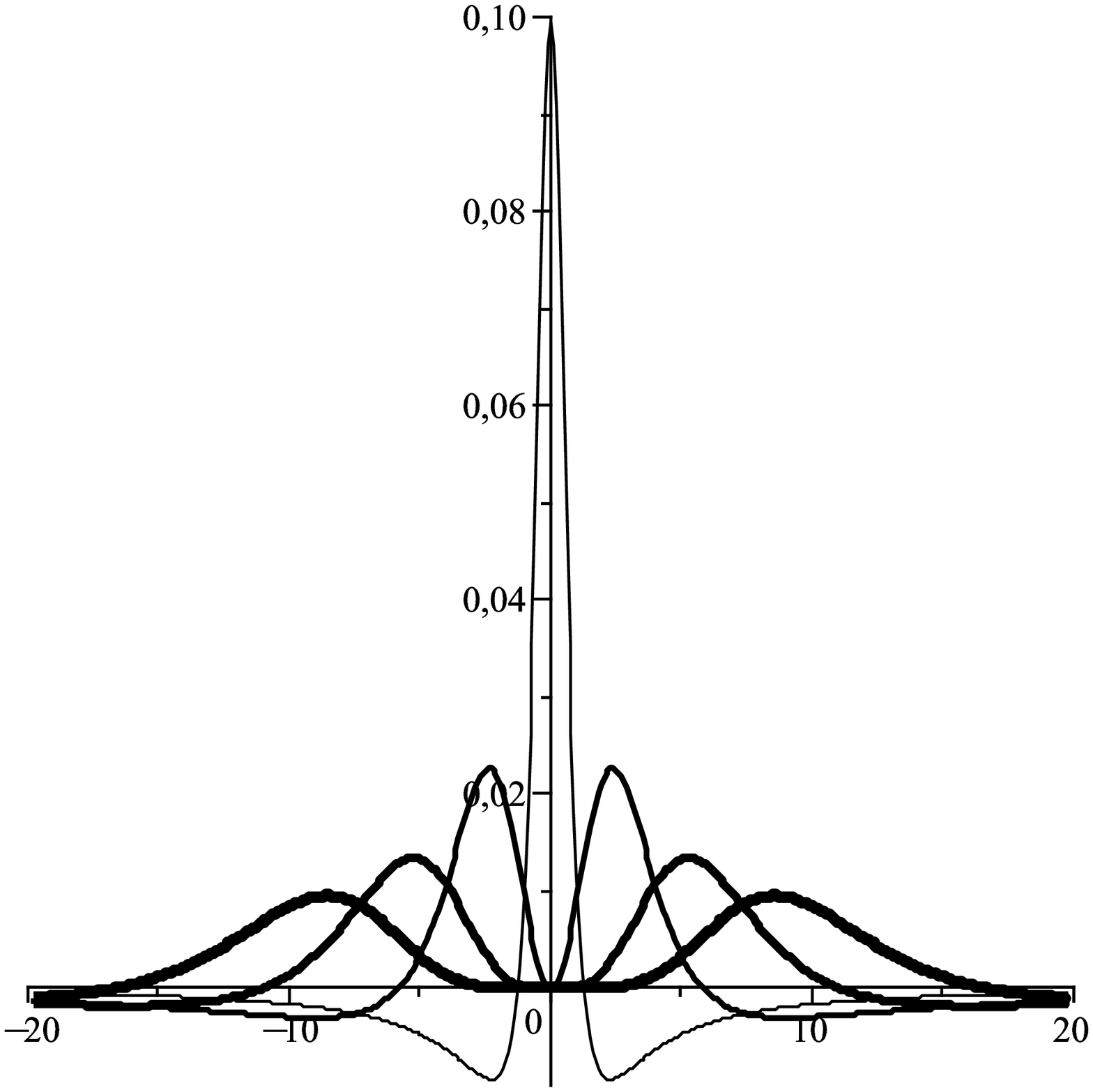}
\includegraphics[width=4.2cm,height=2.5cm]{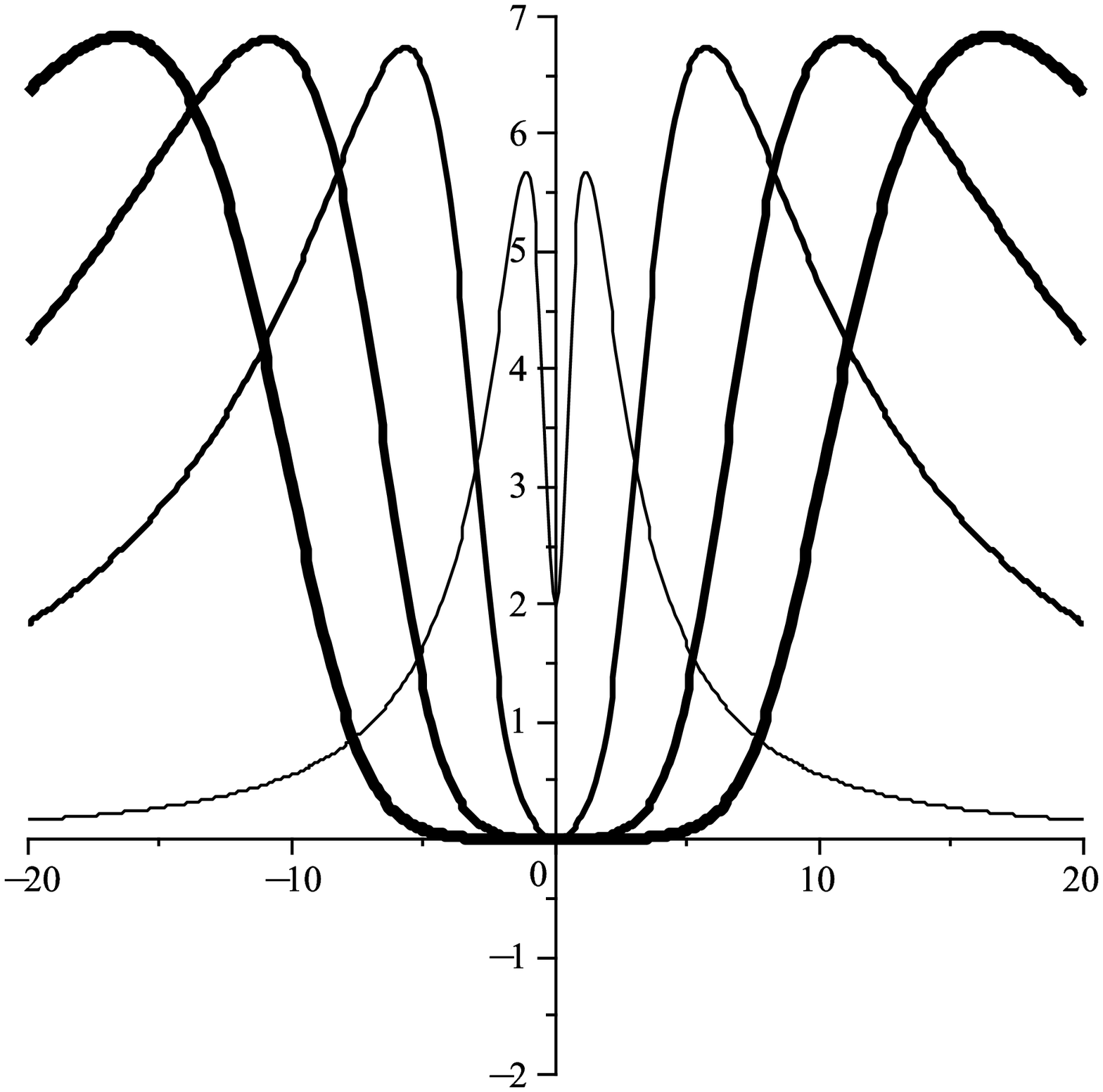}
\includegraphics[width=4.2cm,height=2.5cm]{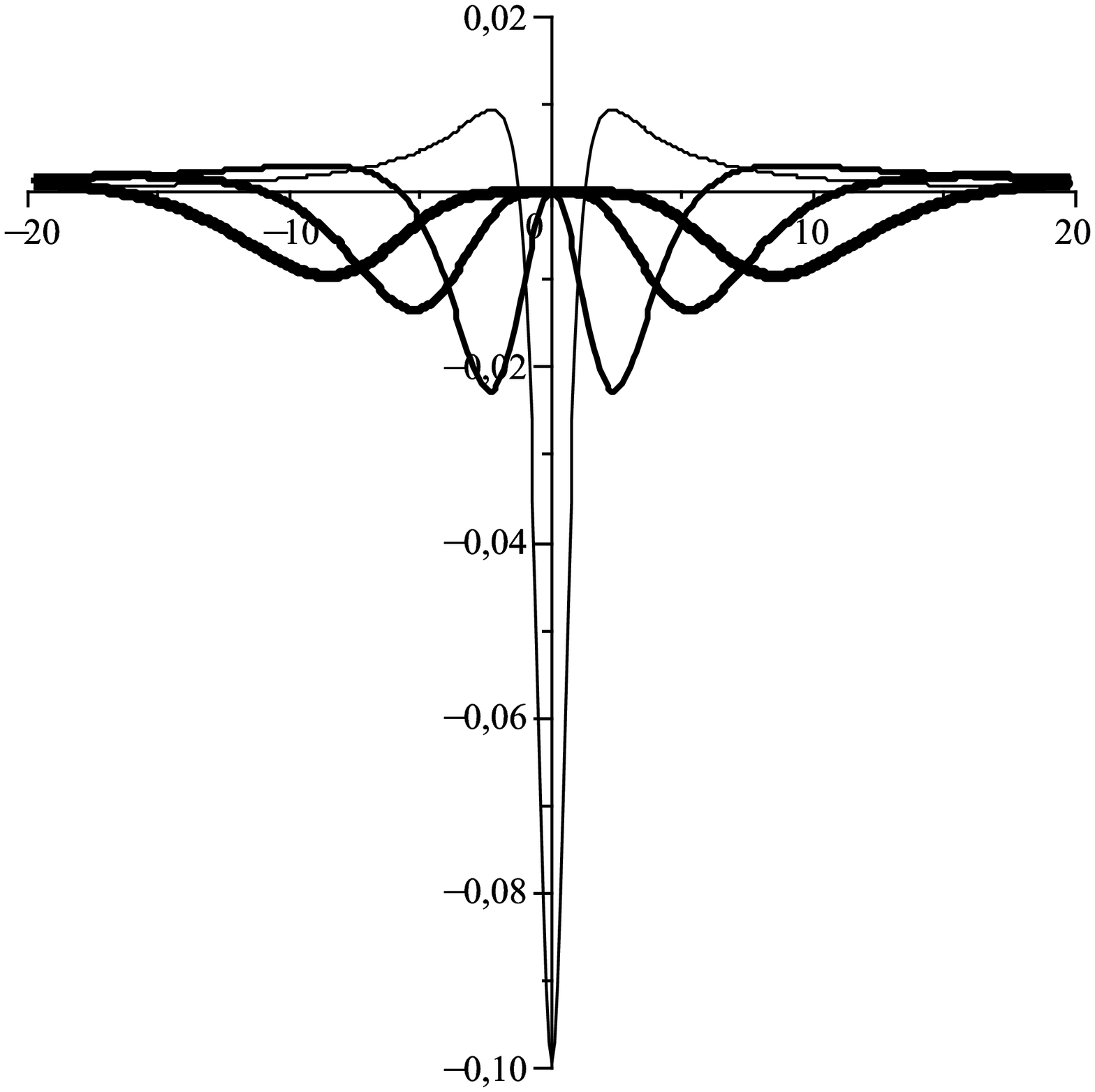}
\includegraphics[width=4.2cm,height=2.5cm]{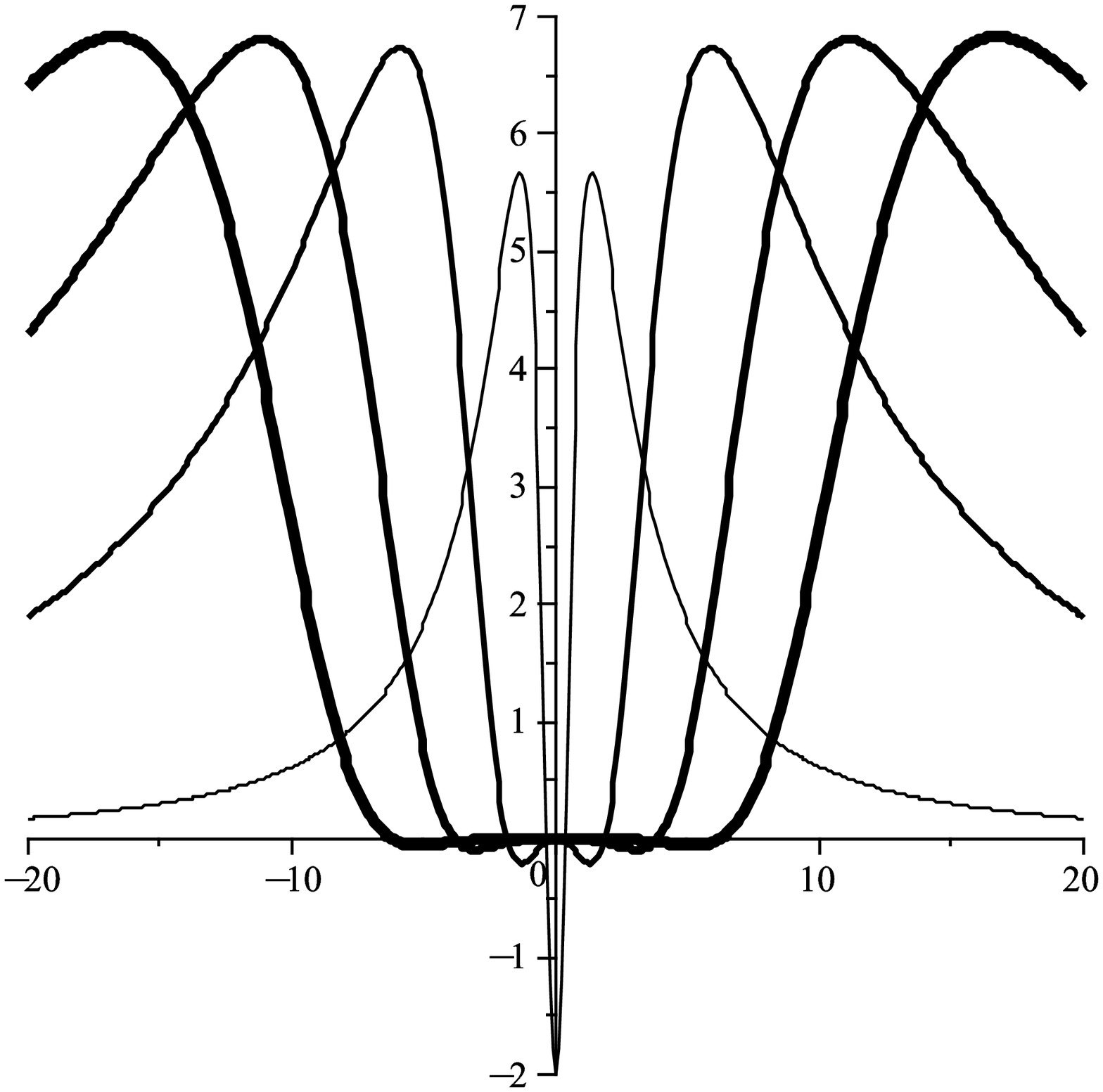}
\caption{\label{neg}Plots of $V_p^R(z)$ (upper) and $V_p^L(z)$ (down)  with (a) $f=0.1$ (left) and (b) $f=2$ (right) fixed. Curves correspond to $p=1$ (tinner line), $3$, $5$ and $7$ (thicker line).}
\end{figure}

In order to investigate the confining of the massive chiral modes, we must solve Eqs. (\ref{schro}) and (\ref{schro2}). It is easy to see that for massive modes that equations can be rewritten as
\begin{small}
\begin{eqnarray}\label{fqm}
\emph{Q}\emph{Q}^+\overline{\alpha}_{Rn}=[\partial_z + f\phi_p e^{A_p}][-\partial_z + f\phi_p
e^{A_p}]\overline{\alpha}_{Rn}=m^2\overline{\alpha}_{Rn},\\\nonumber \emph{Q}^+\emph{Q}\overline{\alpha}_{Ln}=[-\partial_z + f\phi_p e^{A_p}][\partial_z +
f\phi_p e^{A_p}]\overline{\alpha}_{Ln}=m^2\overline{\alpha}_{Ln},
\end{eqnarray}\end{small}
corresponding to a supersymmetric quantum mechanics scenario where tachyonic modes are clearly forbidden.
The changing of variables that produced the Shr\"odinger equations presented in Eq. (\ref{fqm}), lead us to adopt a quantum mechanical interpretation for $\bar\alpha_{Ln}$ and $\bar\alpha_{Rn}$. One important reason for studying resonances
is connected to the information they give for the coupling
between massive modes and the brane. This shows how the
mechanism of fermion trapping is being processed.

In our case, we can interpret   $|N\overline{\alpha}_\pm(z)|^2$  as the probability for finding the massive mode in the position $z$, with $N$ a normalization constant. In this way, calculating $P(m)\equiv|N\overline{\alpha}_\pm(0)|^2$ as a function of the mass $m$, we are able to detect resonant modes as large peaks in the plot of $P(m)$ {\it versus} $m$.

First of all we investigate right chirality, where we already showed there is no zero-mode. Fig. \ref{neg} shows that for the case $p=1$ and small $f$ there is no local minima for the potential $V_p^R(z)$, and resonances are absent. The presence of such minima for $p\ge 3$ shows that resonances possibly exist for a wider range of $f$. In particular for $f=2$, the maxima of the potentials for $p=3,5,7...$ are nearly the same, as well as the local minima. However, the region of the potential near the local minima is broader for larger values of $p$. This signals that for a fixed large value of $f$, larger values of $p$ are more effective in trapping the fermionic massive modes in comparison to smaller values of $p$. This must be compared to the richer structure of the energy density of the branes for larger values of $p$, as noted in \cite{Bazeia:2003aw}.

\begin{figure}
\centering
\includegraphics[angle=-90,width=2.8cm]{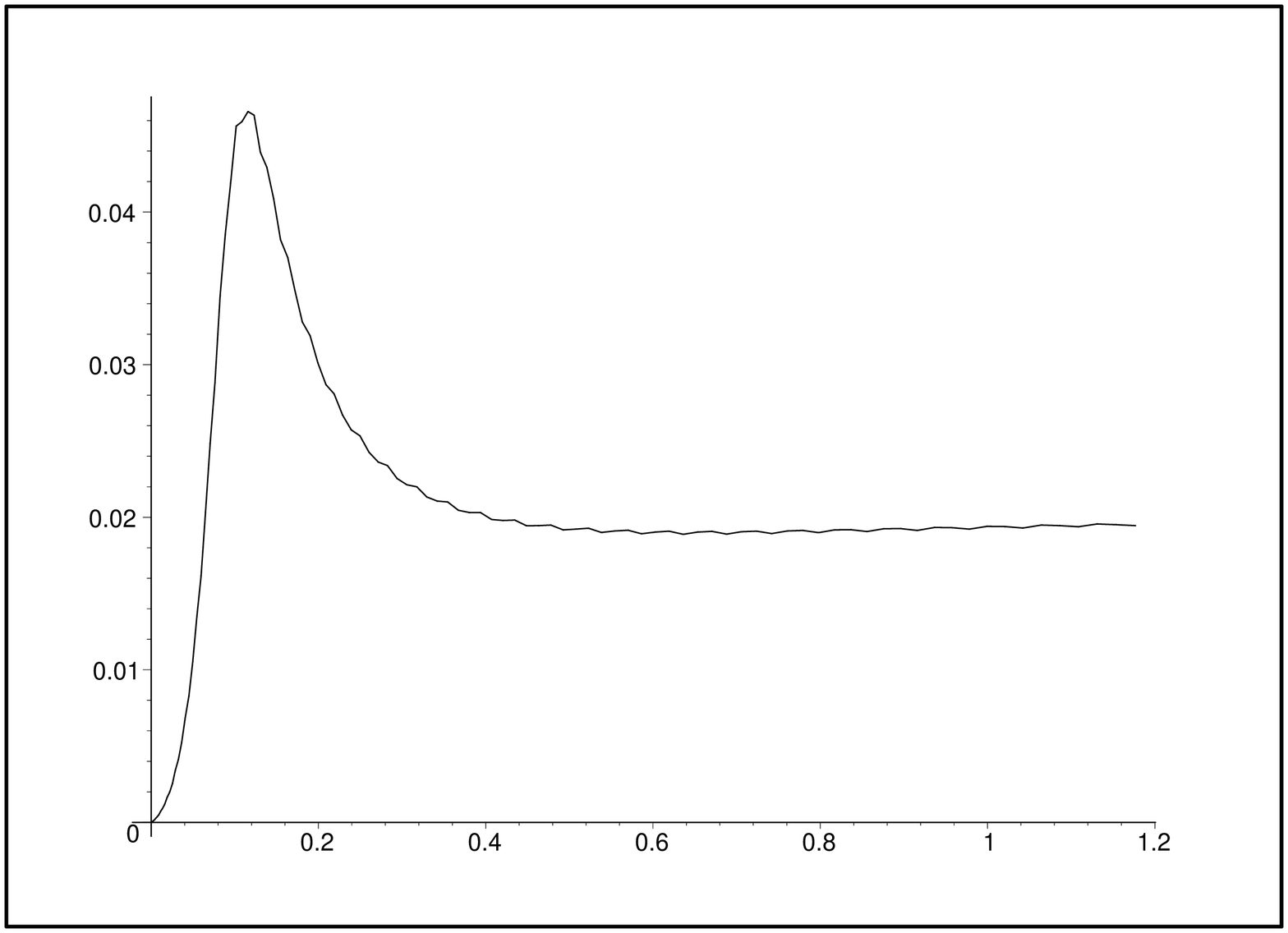}
\includegraphics[angle=-90,width=2.8cm]{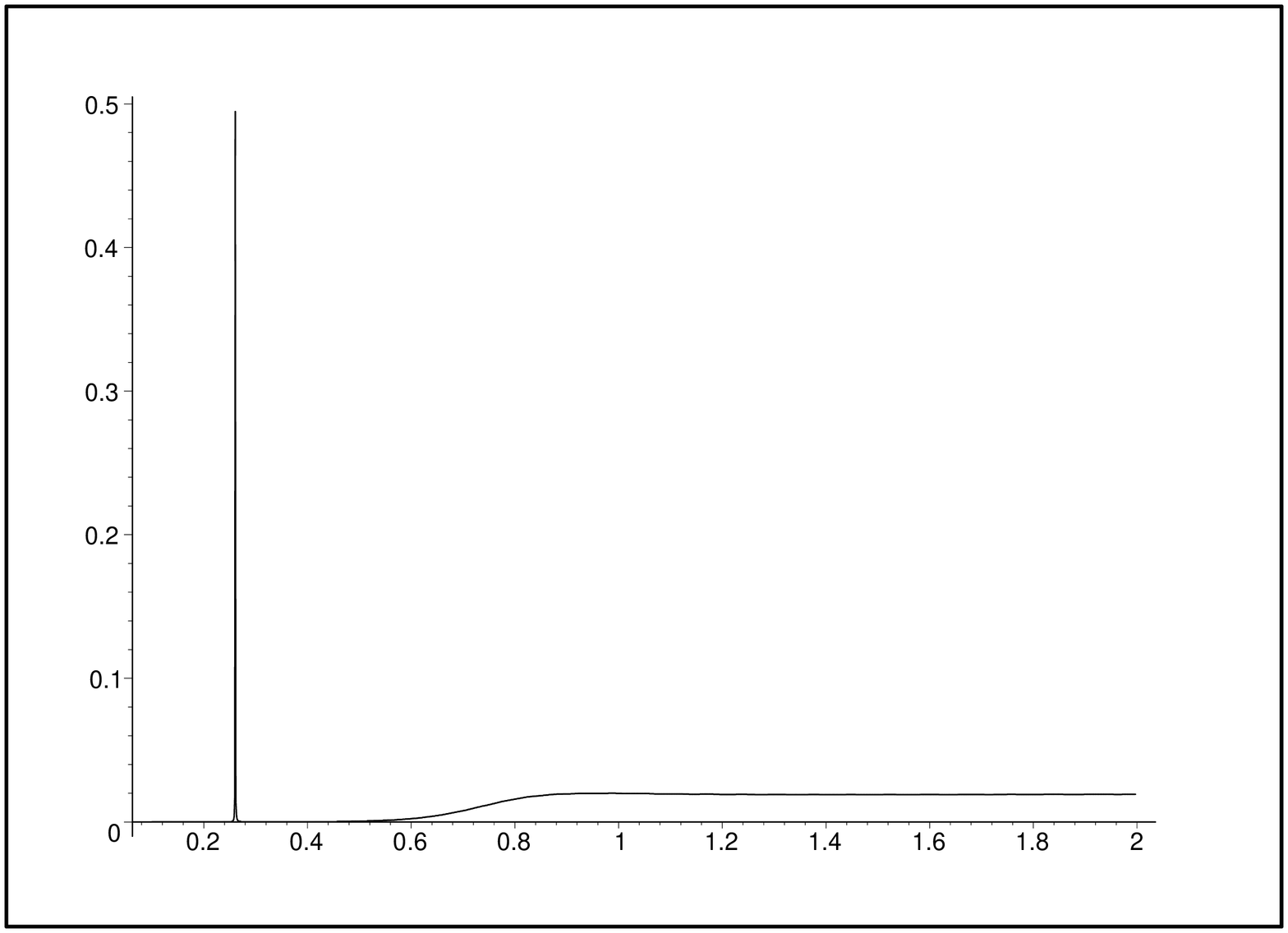}
\includegraphics[angle=-90,width=2.8cm]{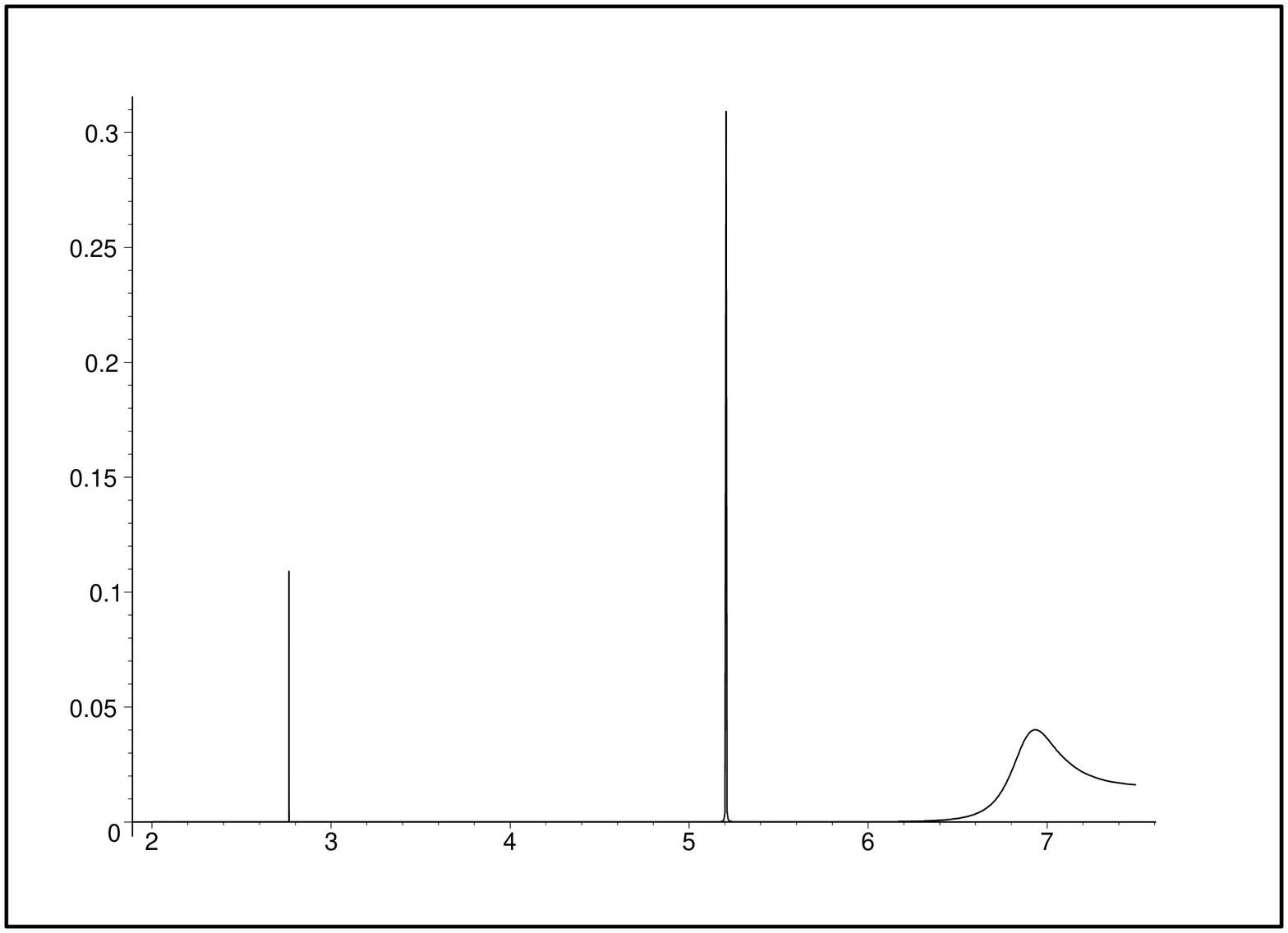}
\includegraphics[angle=-90,width=2.8cm]{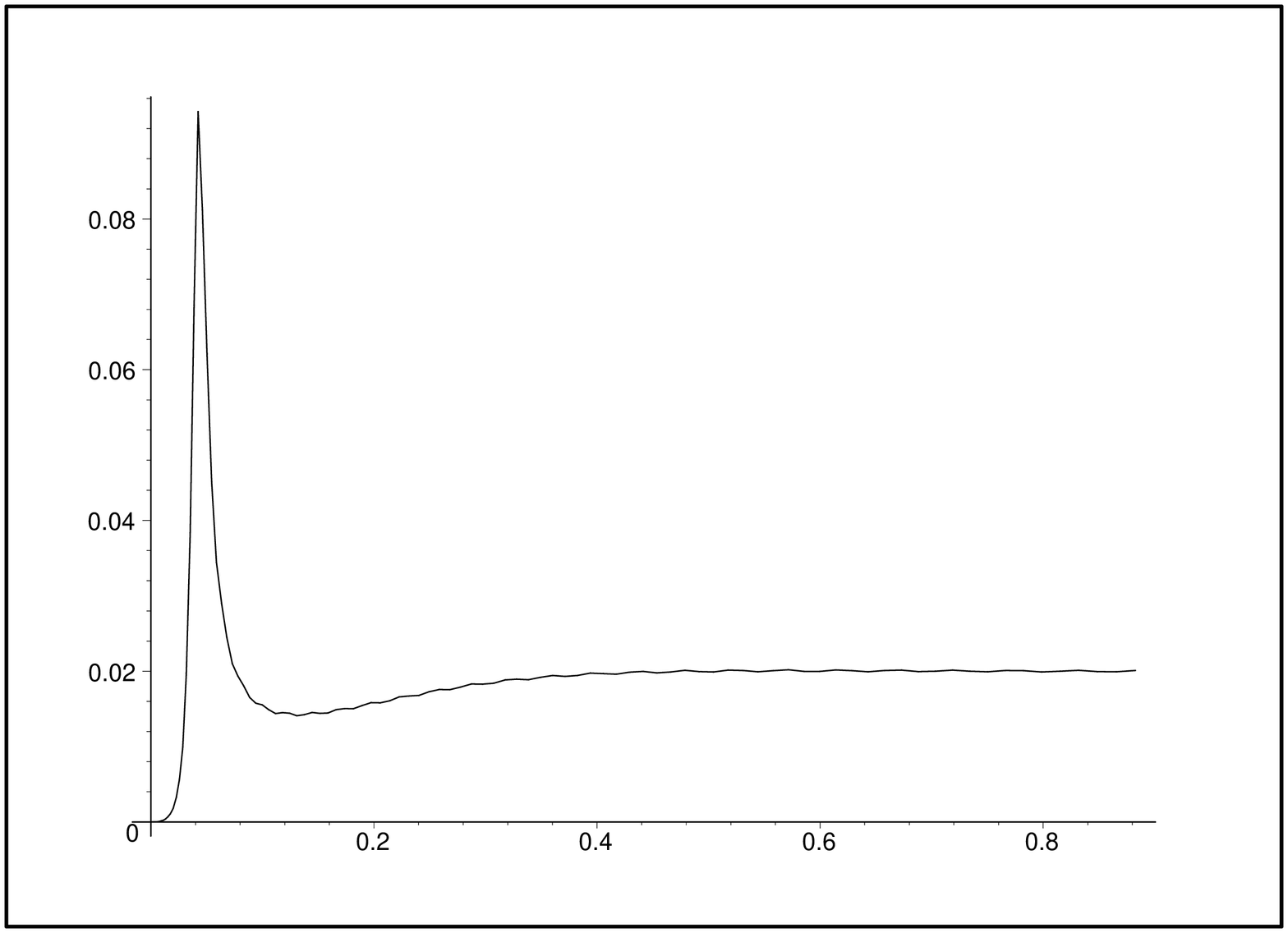}
\includegraphics[angle=-90,width=2.8cm]{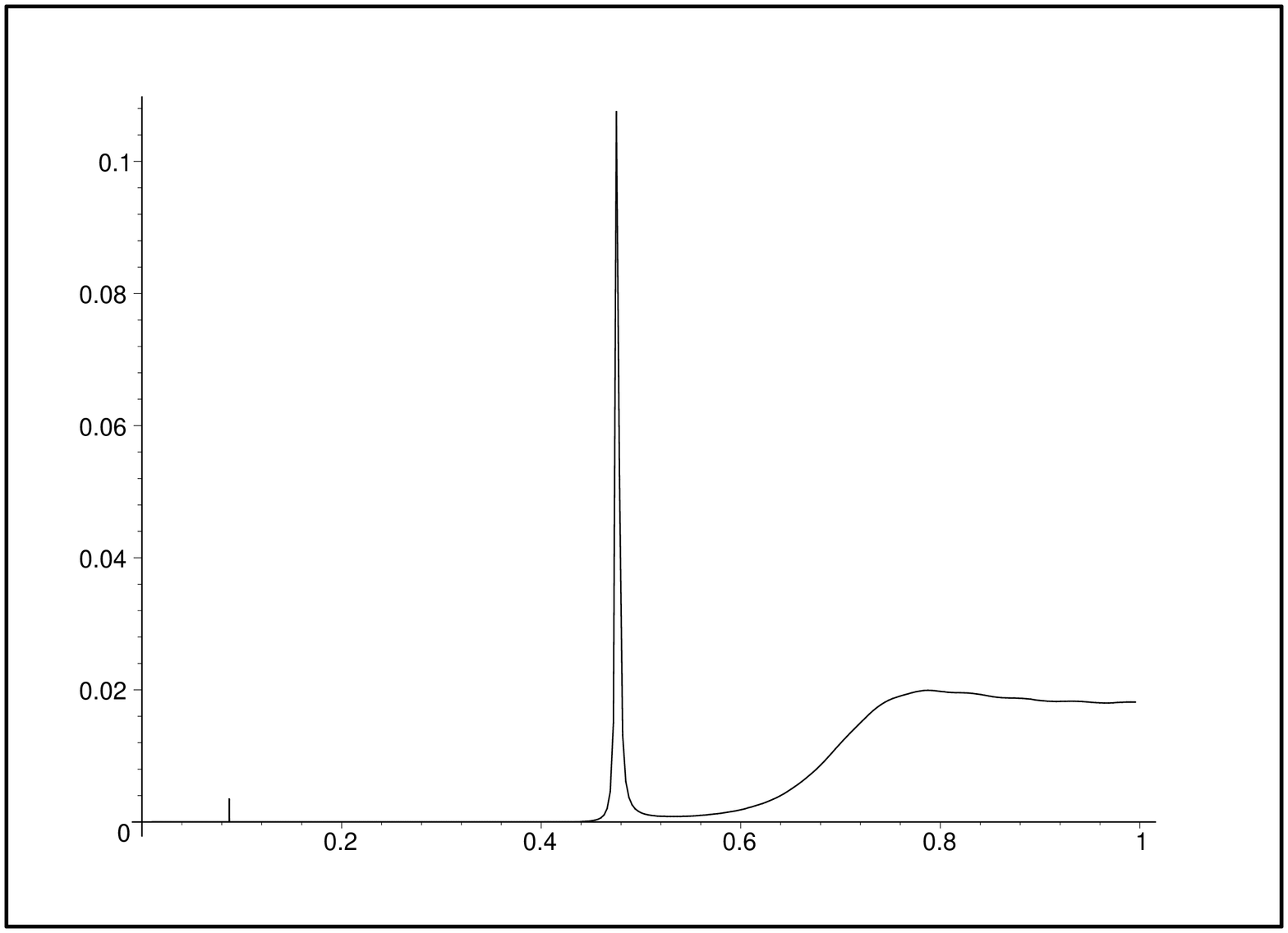}
\includegraphics[angle=-90,width=2.8cm]{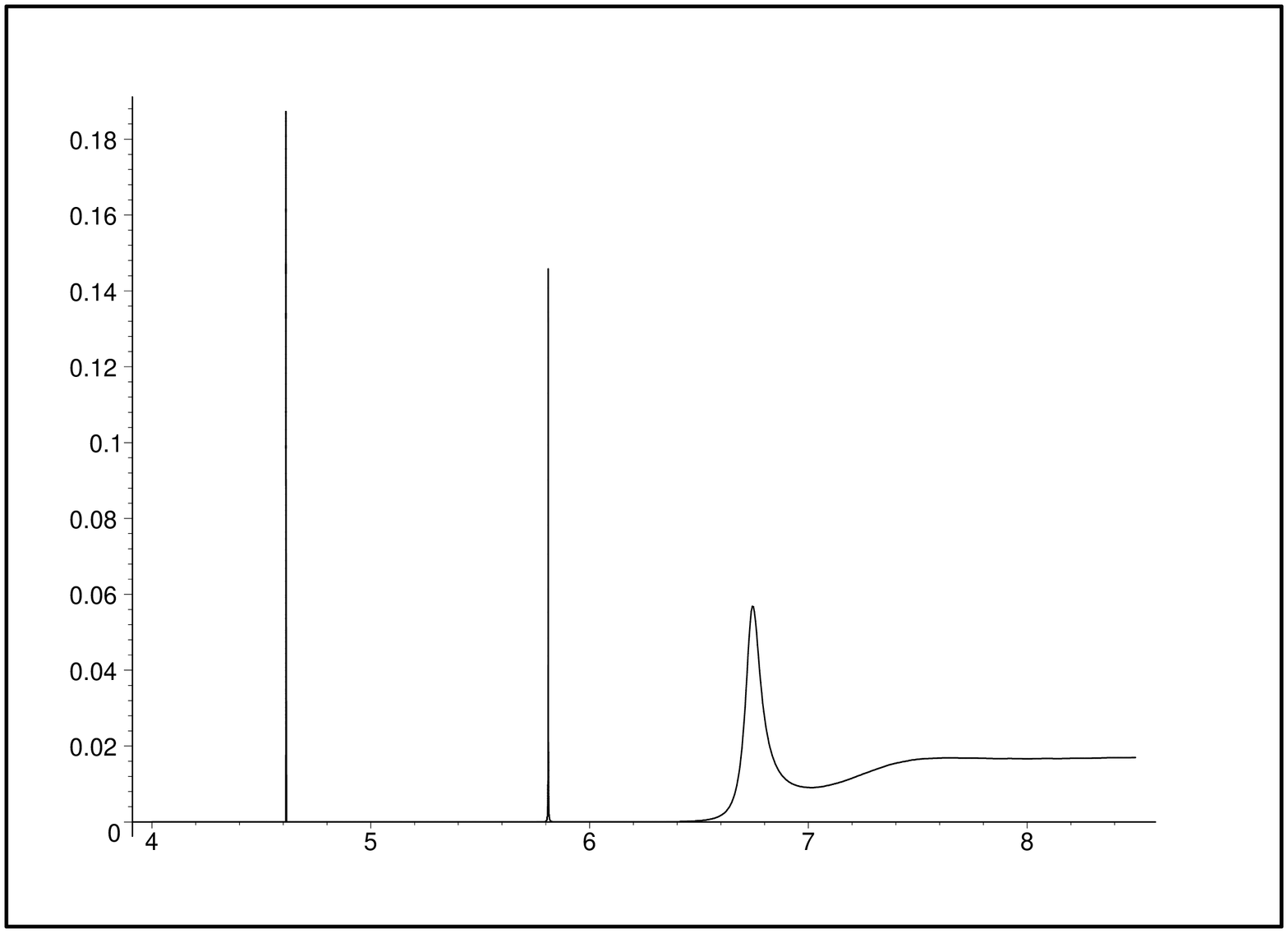}
\includegraphics[angle=-90,width=2.8cm]{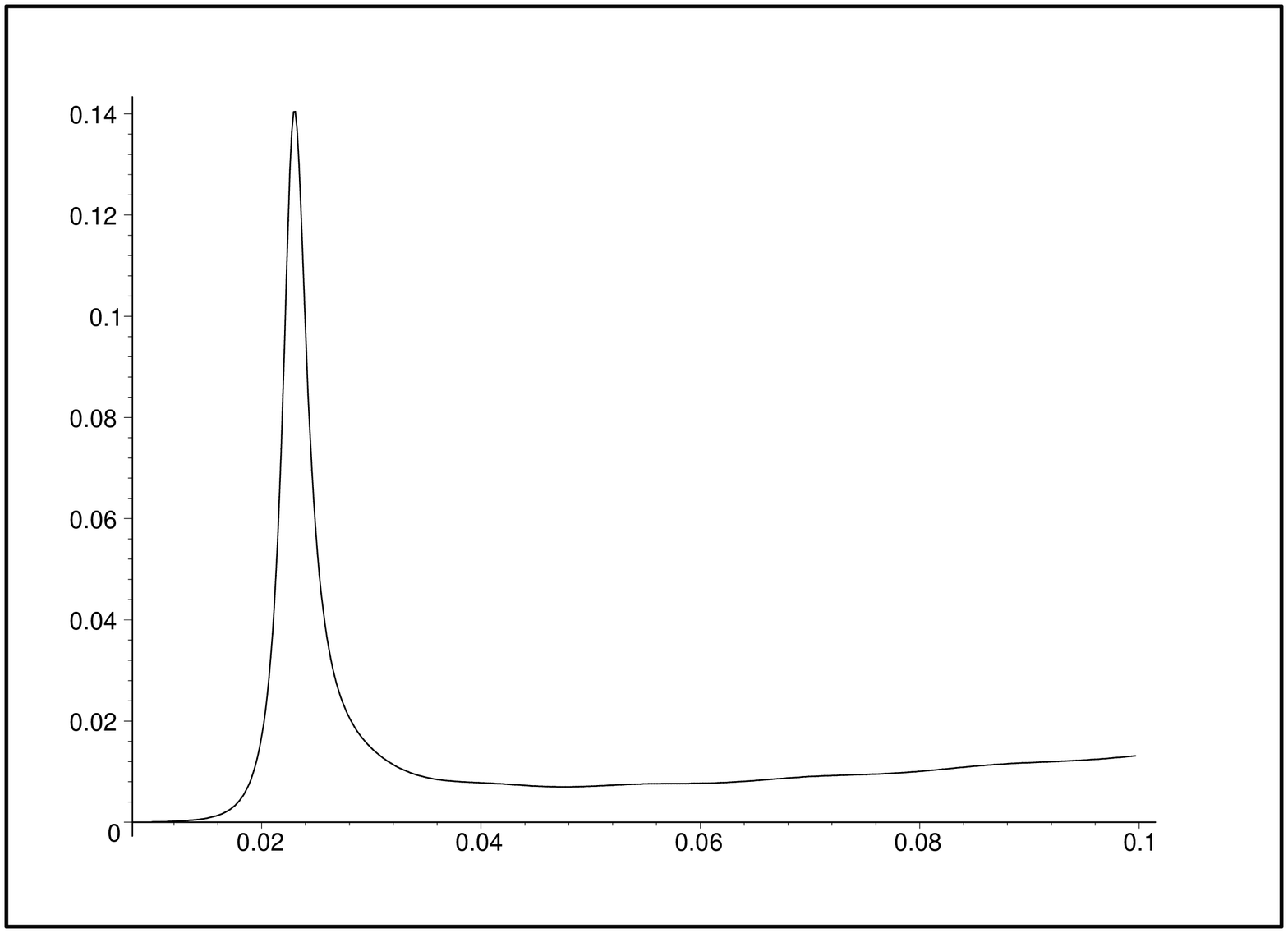}\
\includegraphics[angle=-90,width=2.8cm]{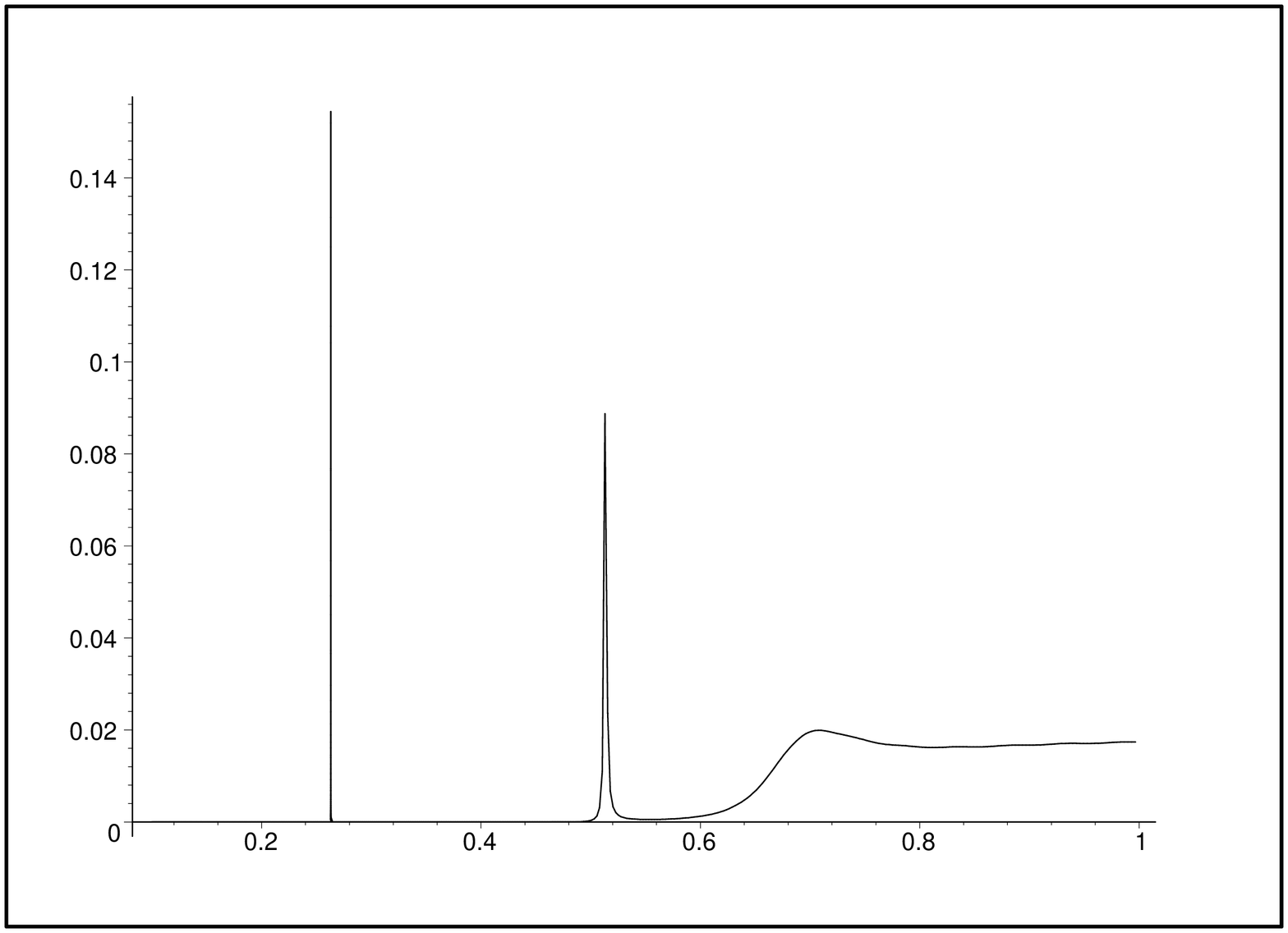}
\includegraphics[angle=-90,width=2.8cm]{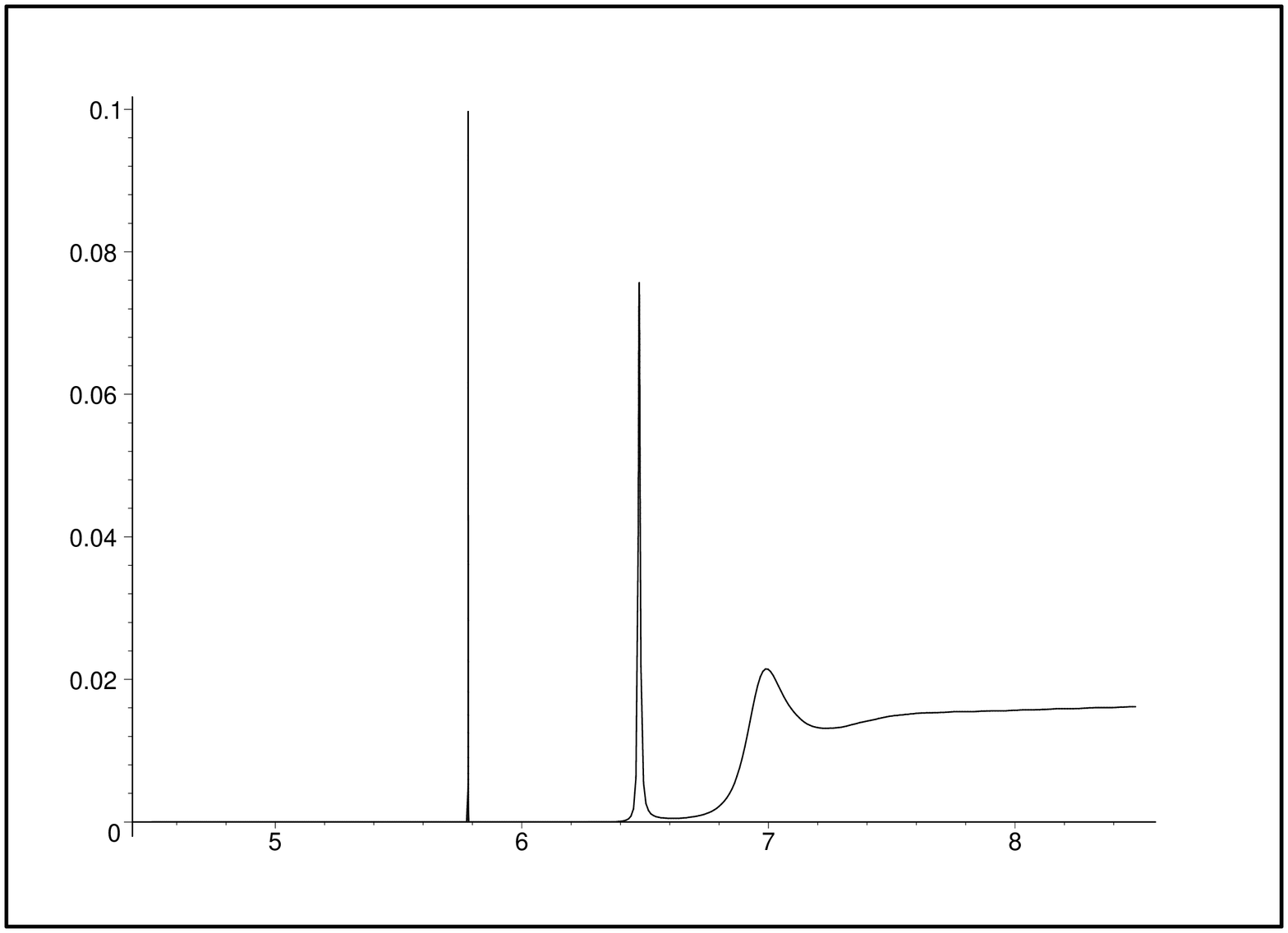}
\caption{\label{probmsq_VR_p3_even} Plots of $|P(0)|^2$ {\it versus} $m^2$ with  $p=3$ (first line), $p=5$ (second line) and $p=7$ (third line). Coupling parameters are $f=0.5$ (left figures), $f=1.1$ (middle figures), and $f=2$ (right figures), for even parity wavefunctions of fermions with right chirality.}
\end{figure}
{\tiny
\begin{table}[htbp]
\centering         \begin{scriptsize}
        \begin{tabular}
        {|l |c | c|r|}\hline {Right} & $f=0.5$ & $f=1.1$ & $f=2$  \\
        \hline $p=3$
 & absent & 0.26061 & 2.764519; 5.2062; 6.94  \\
        \hline $p=5$ & 0.042 & 0.087367; 0.4764 & 4.61154;5.8111; 6.746  \\
        \hline $p=7$
 & 0.0230 & 0.263018; 0.5134 & 4.9906;5.7828;6.476;6.9910 \\
        \hline{Left}  & $f=0.5$ & $f=1.1$ & $f=2$  \\
        \hline $p=3$
 & absent & 0.646 &  4.63126; 6.593 \\
        \hline $p=5$ & absent & 0.29816; 0.640 &  5.52762; 6.552 \\
        \hline $p=7$
 & 0.0432 & 0.623; 0.40240& 5.59193; 6.318; 6.872\\
        \hline
        \end{tabular}
               \caption{\it First resonance peaks. The table shows the corresponding values of $m^2$.}  \end{scriptsize}
   \end{table}}

We found resonance peaks for even parity wavefunctions and analyzed the influence of the parameter $p$ and coupling constant $f$. Some results for right chiral fermions are shown in Fig. \ref{probmsq_VR_p3_even} for $p=3, 5$ and $7$ and three values of $f$.  The resonances found are shown in Table I.

For $p=3$ and $f=0.5$ we see that there is no resonance, as the width at half maximum $\Delta m$ of the peak is bigger than the mass $m$ corresponding to the peak; for $f=1.1$ and $f=2$ there appears one and three resonances, respectively. This agrees with our expectation that larger values of $f$ favor the presence of resonances. Note that in general, for fixed $p$, larger values of $f$ lead to larger number of peaks. Also, for fixed $f$, larger values of $p$ corresponds to the thicker peaks with almost the same mass. The first peak is the thinnest, with corresponding larger lifetime. The increasing of the parameter $p$ tends to increase the mass of the first resonance (this is more evident for larger values of $f$), whereas the increasing of $f$ turns richer the spectrum, with more resonances.

\begin{figure}
\centering
\includegraphics[angle=-90,width=2.8cm]{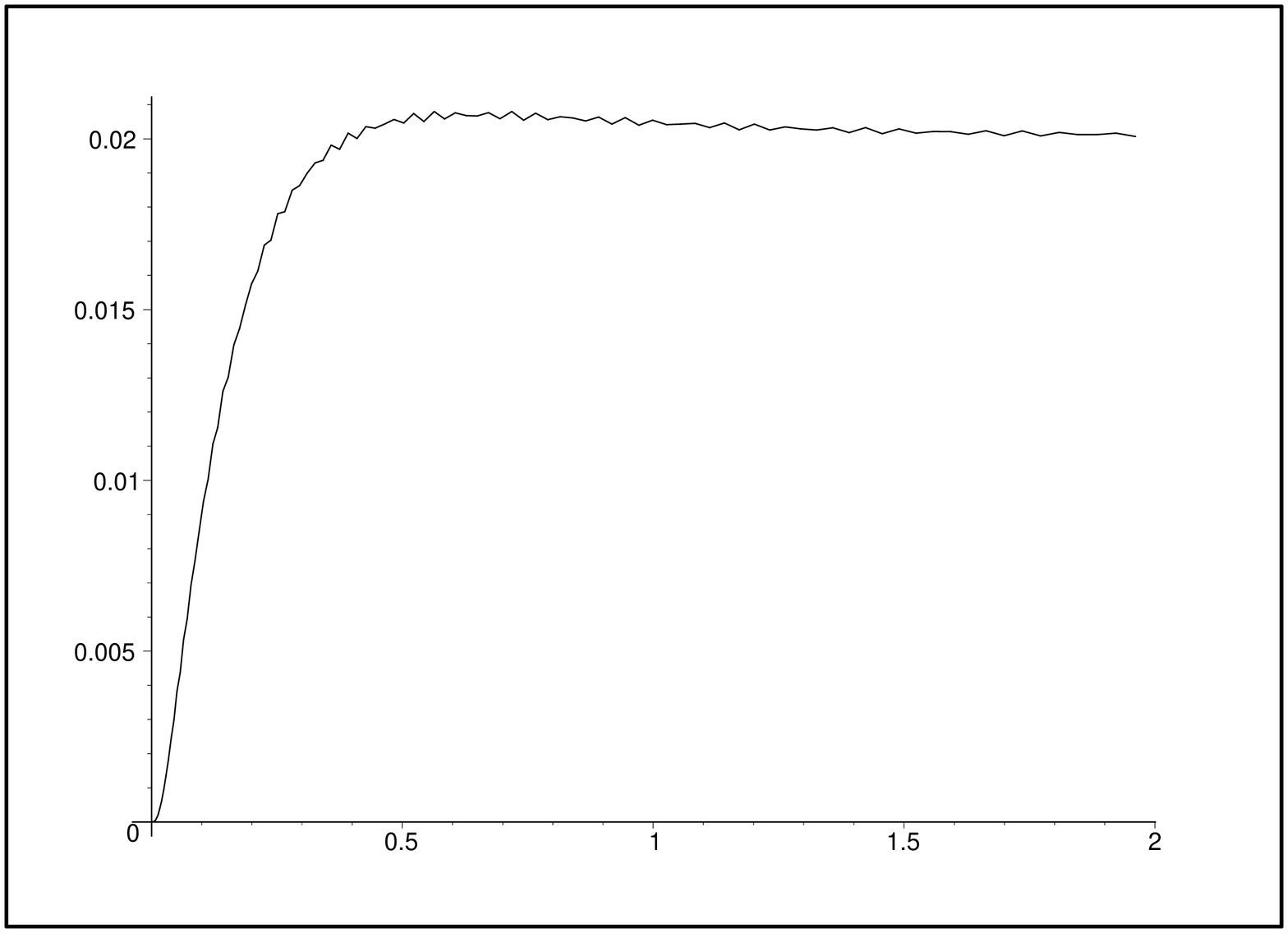}
\includegraphics[angle=-90,width=2.8cm]{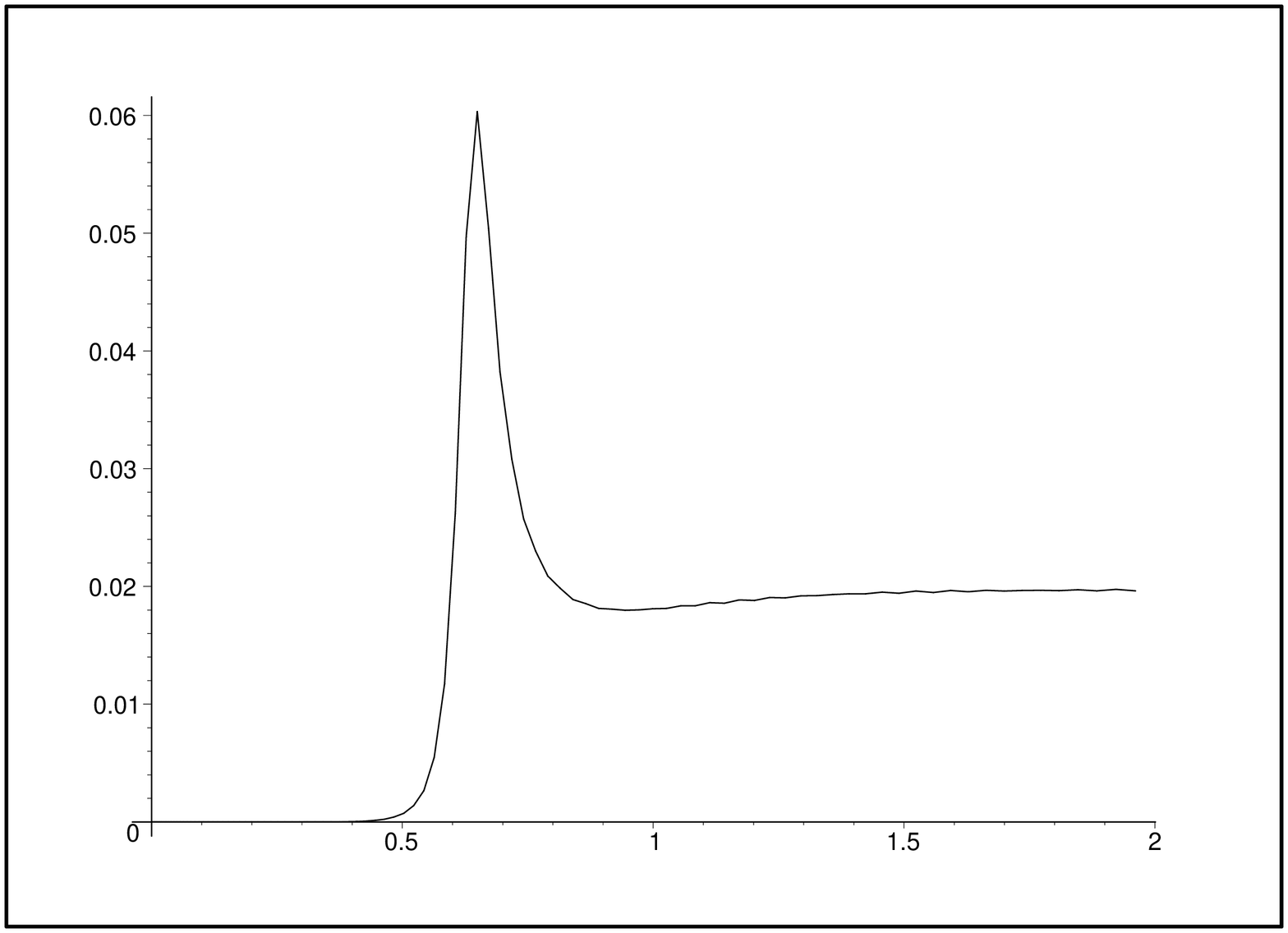}
\includegraphics[angle=-90,width=2.8cm]{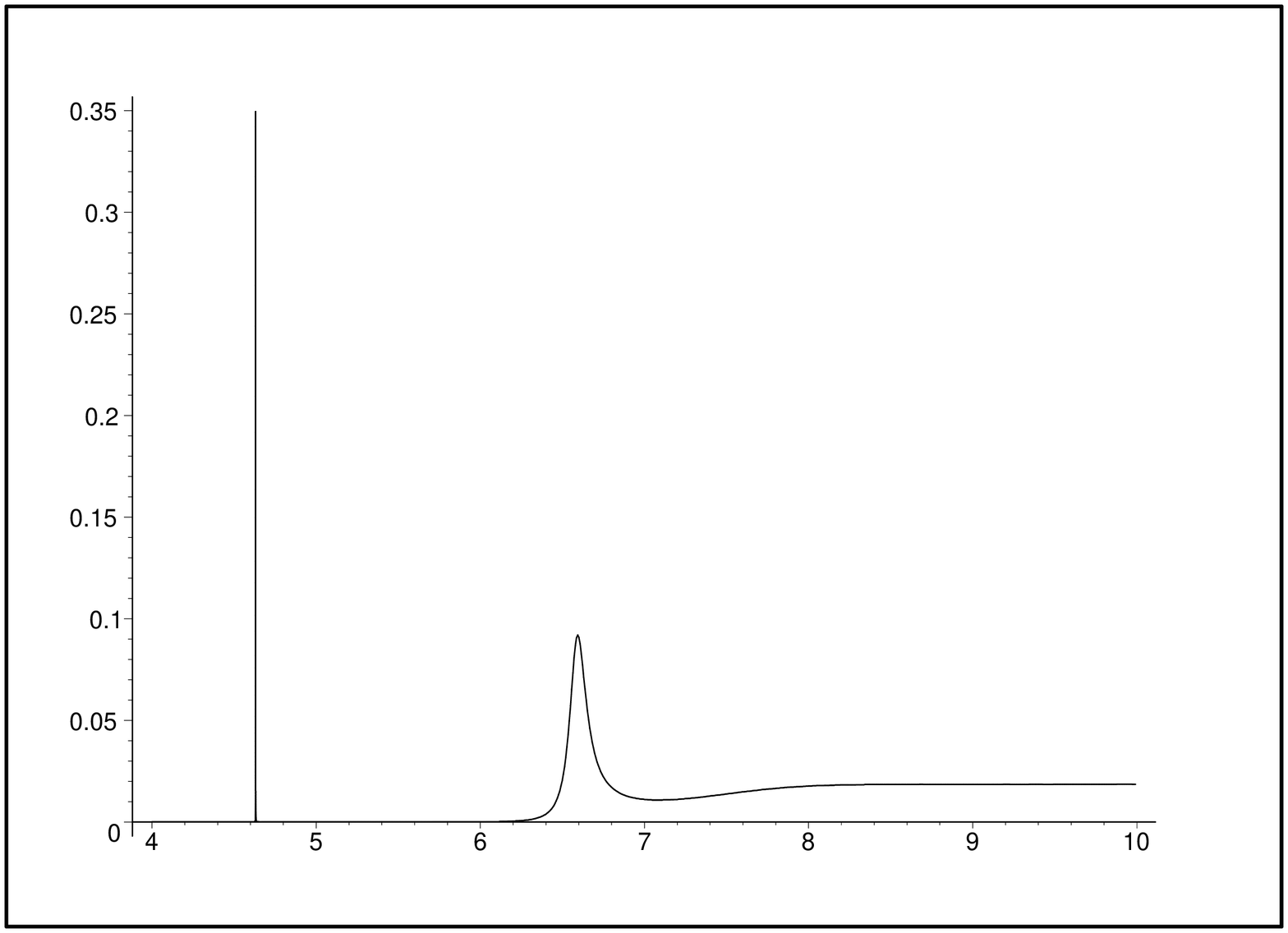}
\includegraphics[angle=-90,width=2.8cm]{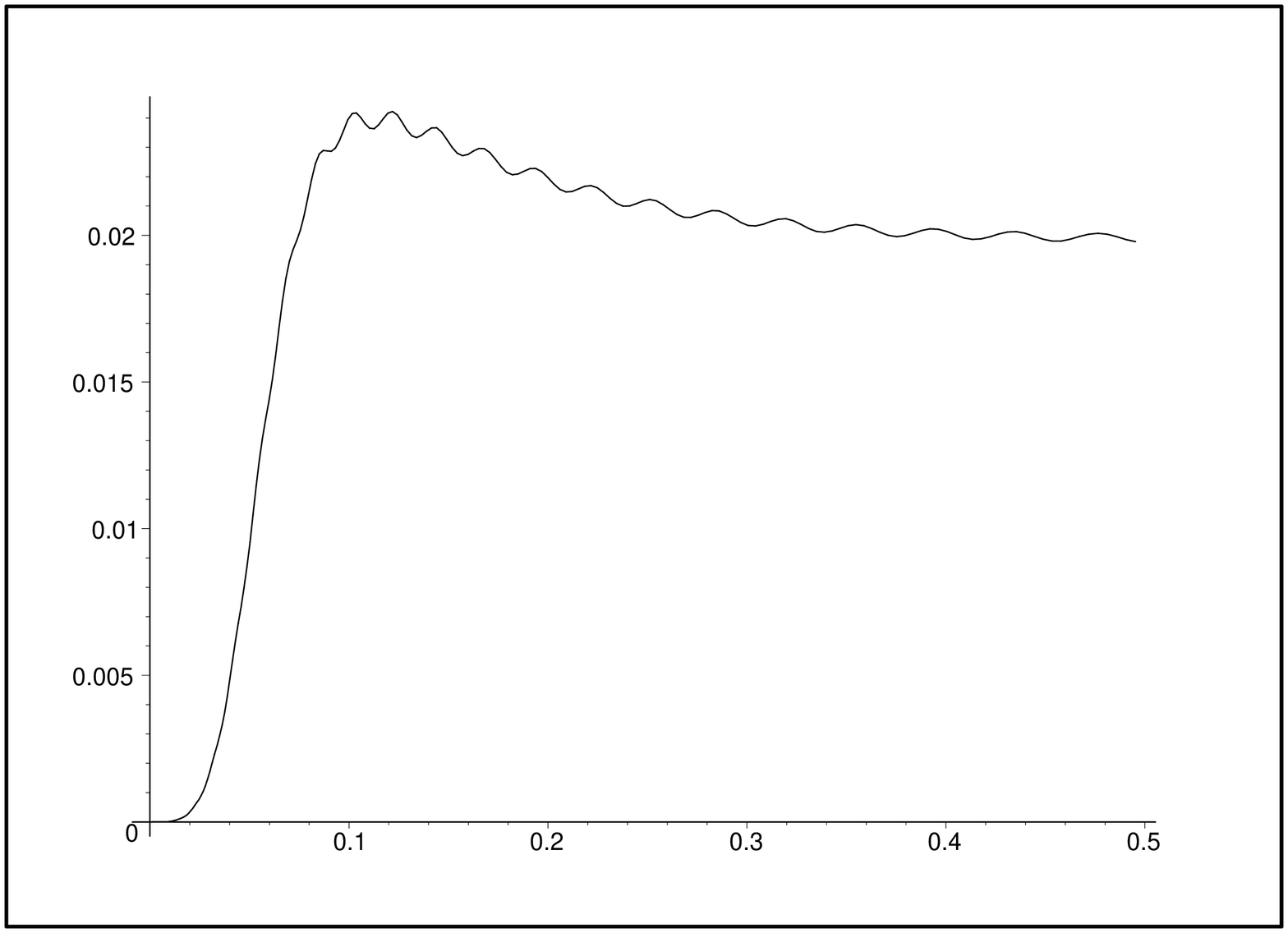}
\includegraphics[angle=-90,width=2.8cm]{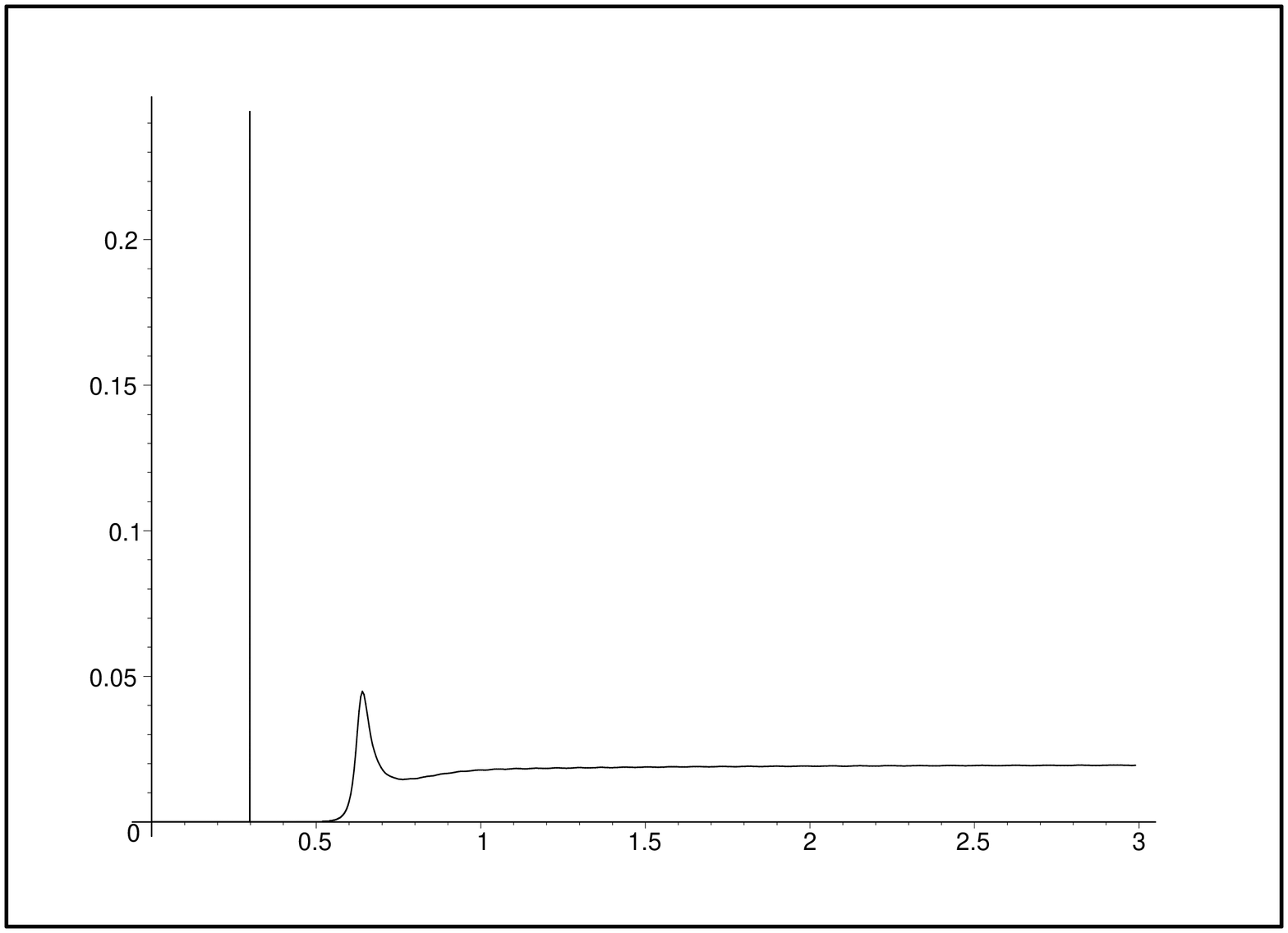}
\includegraphics[angle=-90,width=2.8cm]{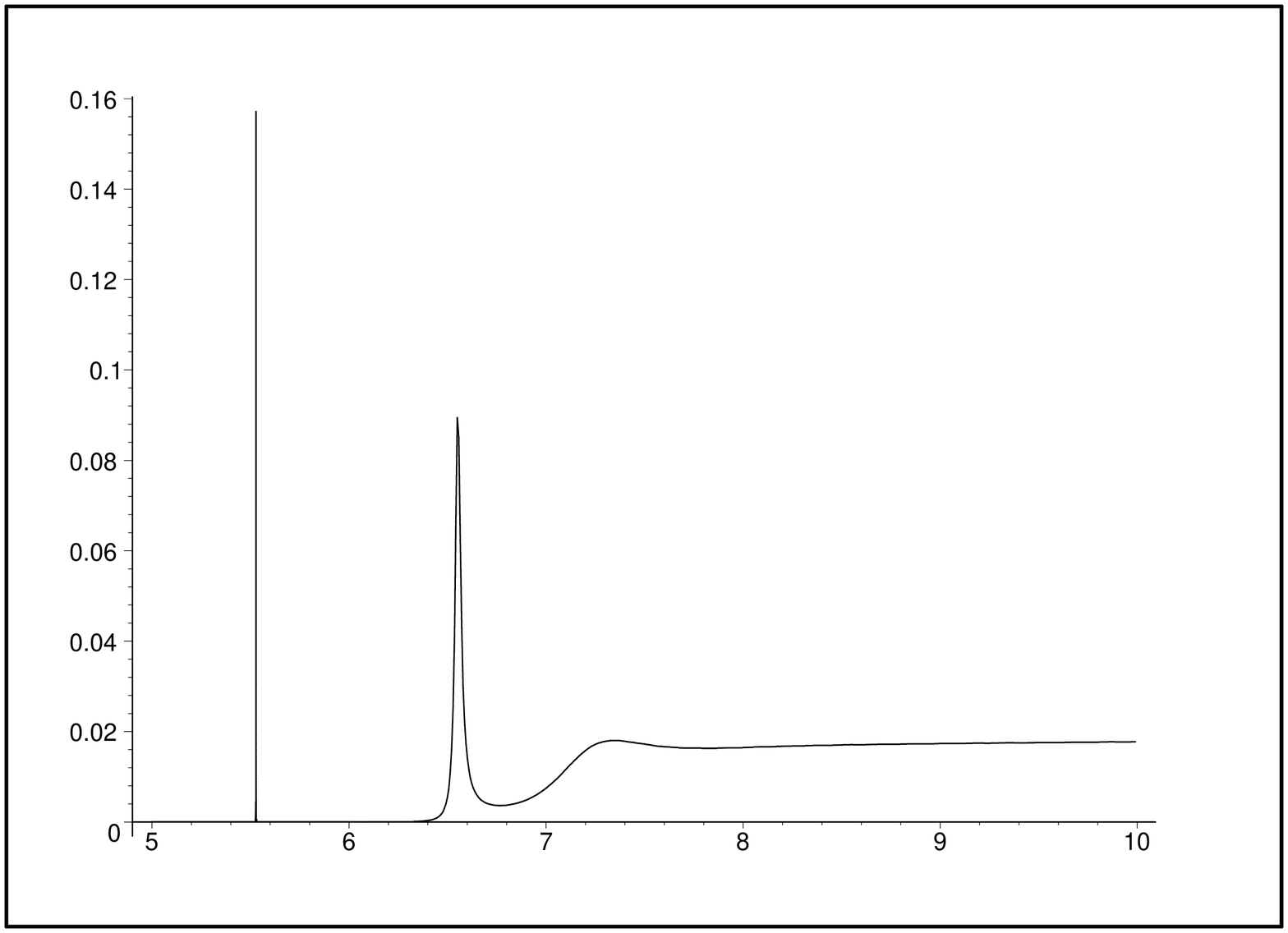}
\includegraphics[angle=-90,width=2.8cm]{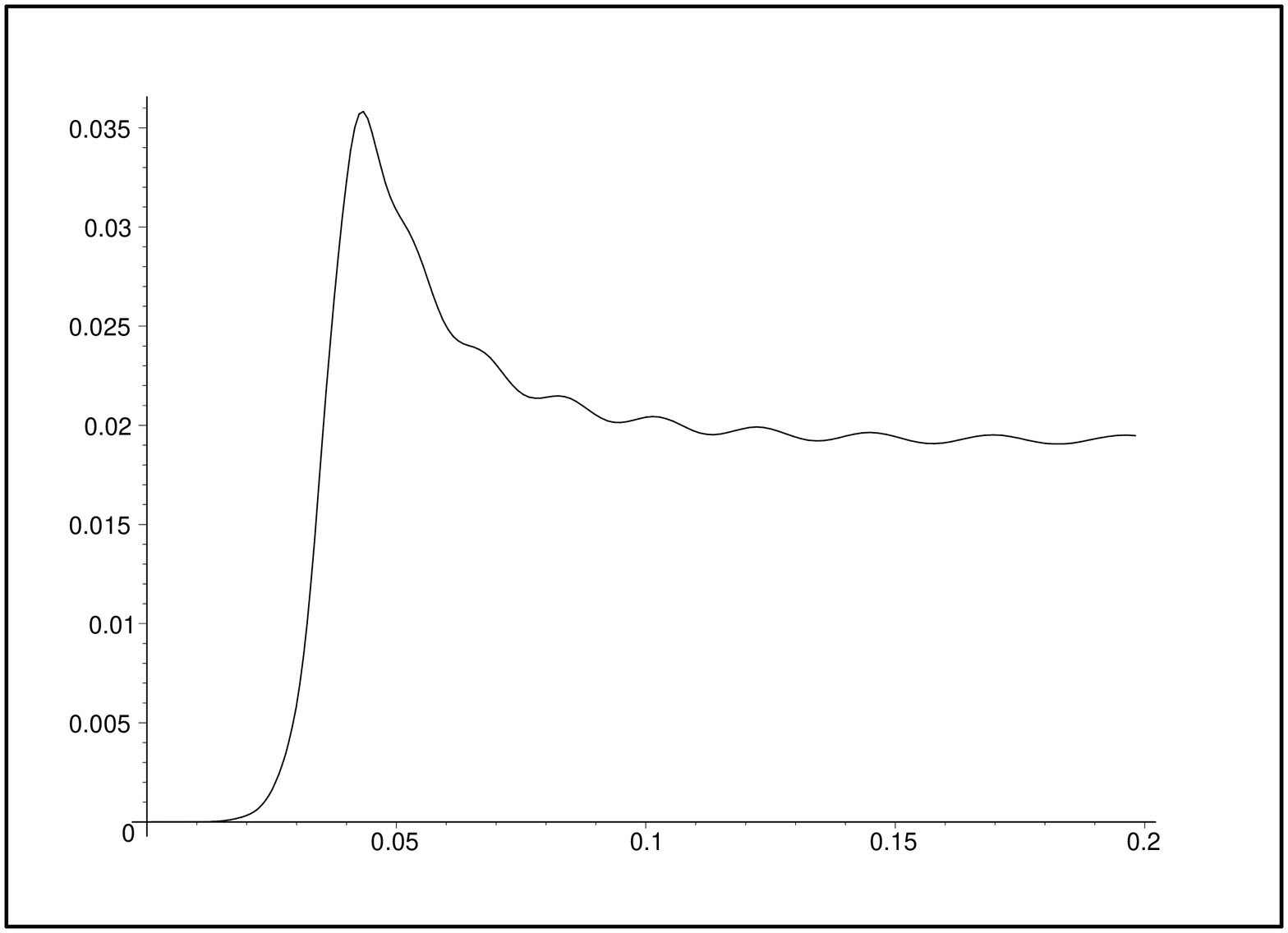}
\includegraphics[angle=-90,width=2.8cm]{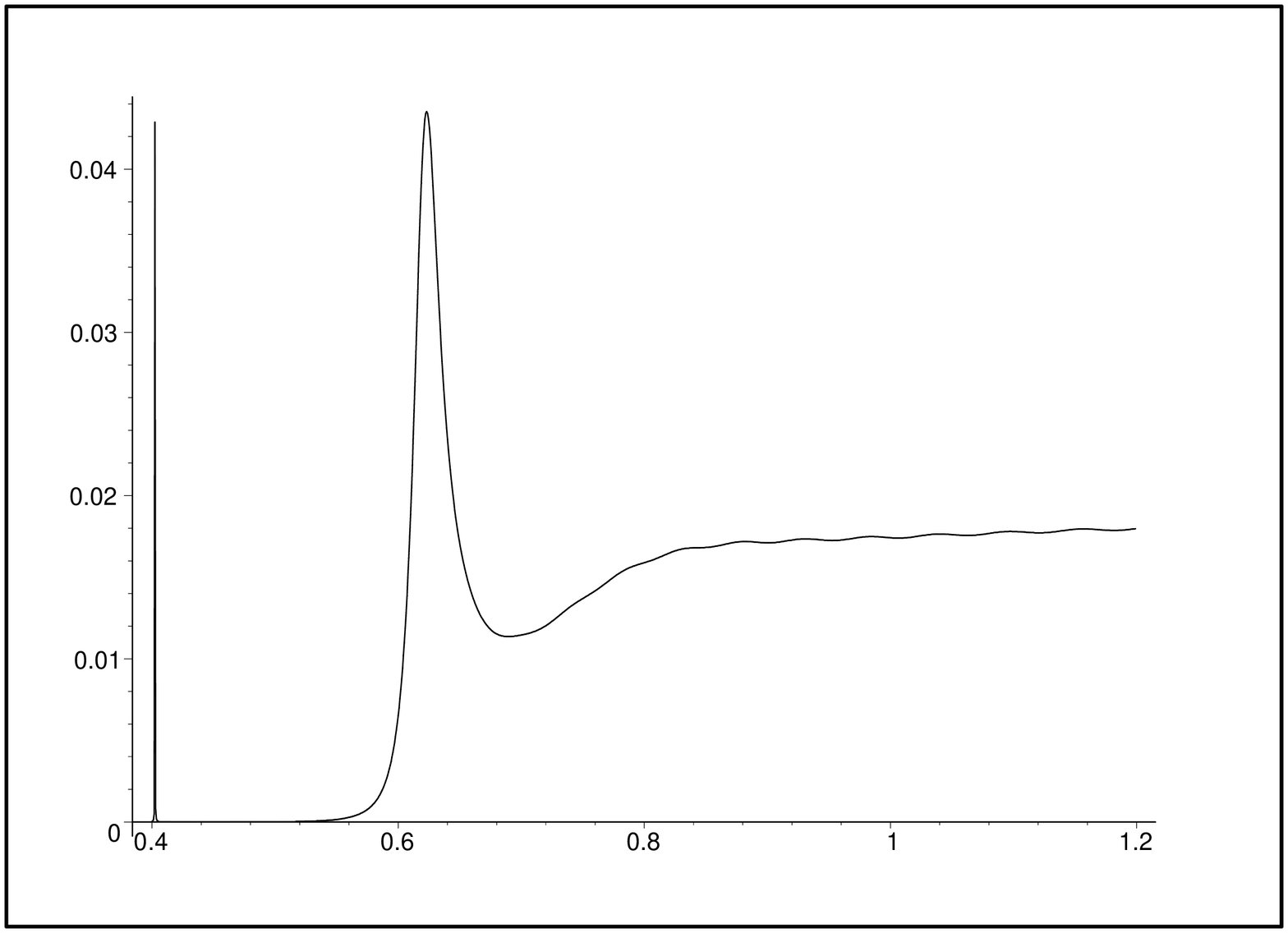}
\includegraphics[angle=-90,width=2.8cm]{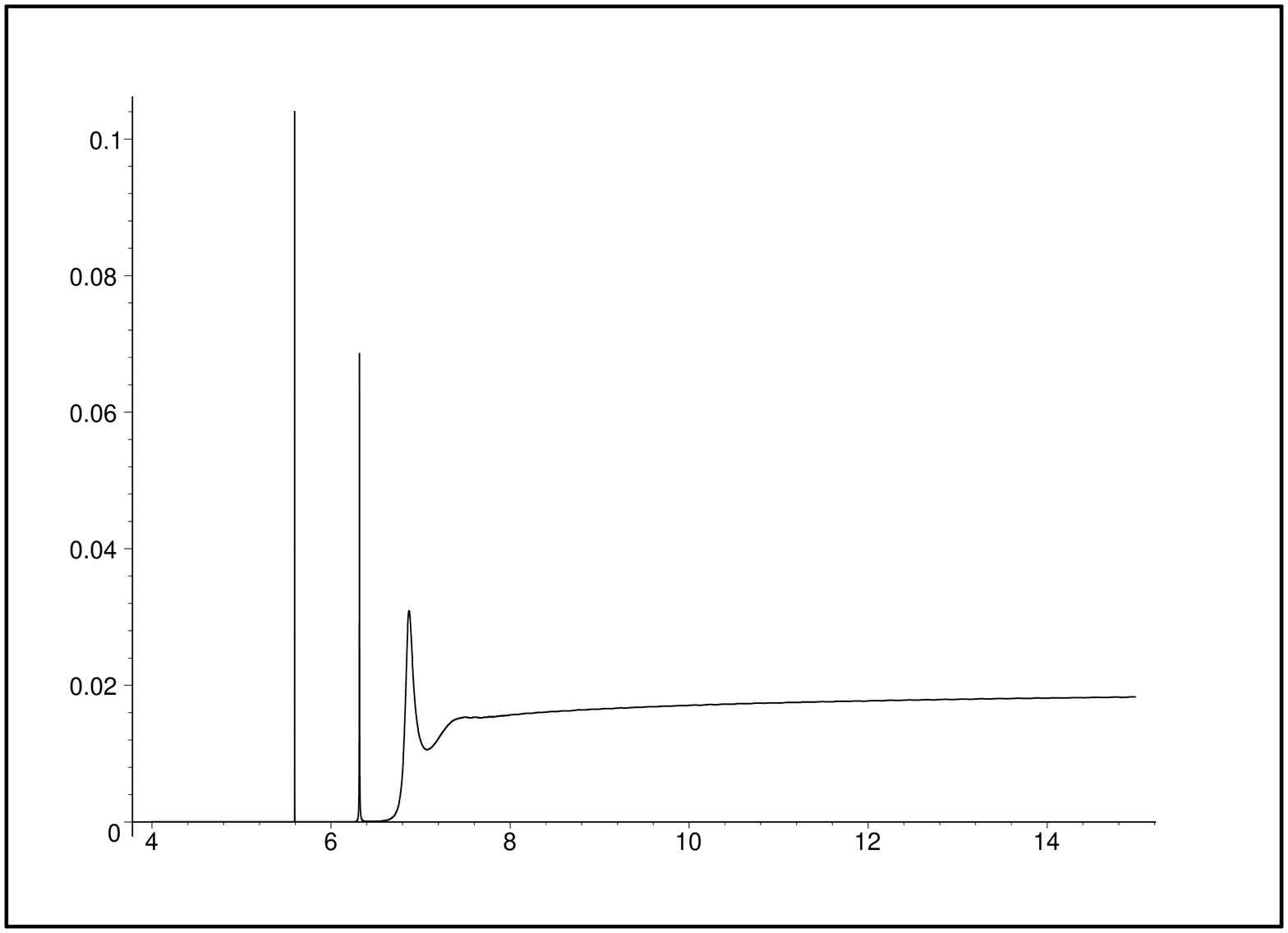}
\caption{\label{probmsq_VL_p3_even} Plots of $|P(0)|^2$ {\it versus} $m^2$ with  $p=3$ (first line), $p=5$ (second line) and $p=7$ (third line). Coupling parameters are $f=0.5$ (left figures), $f=1.1$ (middle figures), and $f=2$ (right figures), for even parity wavefunctions of fermions with left chirality.}
\end{figure}

Next we analyzed left chiral fermions. Fig. \ref{probmsq_VL_p3_even} shows the resonance peaks for $p=3,5,7$ with same values of $f$ used previously, but now for left chirality.
We found the same effects related to the influence of $p$ and $f$. In particular, note that the mass corresponding to the broader resonance peak did not change significantly with $p$. One remarkable point observed with the increase with $p$ is the tendency for the resonances corresponding to tiny peaks to accumulate near the broader peak.
\begin{figure}
\centering
\includegraphics[angle=-90,width=6cm]{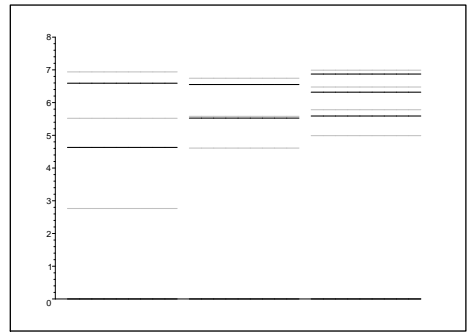}
    \caption{\label{spectrum} Spectrum for $p=3$ (left),  $p=5$ (middle) and $p=7$ (right) with $f=2$ fixed, constructed with even parity modes. Black lines for left chiral massive modes. Grey lines are for right chiral ones.}
\end{figure}

Now we summarize our findings for the resonance spectrum for both chiralities. First of all note that from Eqs. (\ref{fqm}) the spectrum of the right and left chiralities are related. Indeed, the spectrum starts with zero mode with right chirality and even parity. The two resonances with same $m^2$ that succeed in the spectrum are the first odd parity left chiral mode and the first even parity right chiral mode. Next we also have two resonances with same $m^2$: the first even parity left chiral mode and the first odd parity right chiral mode.

We successively have in the spectrum even and odd parity wavefunctions for left and right chiral modes with same values of $m^2$. One can easily check the formation of Dirac fermions after studying the odd parity wavefunctions. This needs the changing the normalization procedure in a known procedure \cite{Liu2} (for an analysis with models with two scalar fields, see Ref. \cite{ca}).We checked the correspondence of the spectrum for some values for resonances with odd parity wavefunctions, but we will not pursue this subject in detail, since the numerical procedure was tested with confidence and the formation of the Dirac fermions are guaranteed by the supersymmetric quantum mechanics structure.

The spectra for $p=3,5,7$ and $f=2$ are depicted in Fig. \ref{spectrum}. There one can see how the increasing in $p$ increases the number of resonances. Note also that the resonances appear coupled in pairs for left and right chiralities (here we consider only even parity wavefunctions). We also see that an increasing of $p$ increases the mass of the first resonance, generally being a very thin line in the spectrum. The fact that the lines for left and right chiralities alternate each other in the spectrum is very important for capture the thinner lines in the numerical procedure.

\section{Conclusions}
In this work we have first studied localization of fermionic zero-modes after introducing a Yukawa coupling between the 5-dimensional spinor and the scalar field, depending on a parameter $f$. A significant result was that the parameter that controls the class of deformation has established the condition for finding zero modes. This condition was observed by a direct relation between the Yukawa coupling constant and the  deformation parameter. We found that zero-mode exists only for left chirality in a specific relation between the Yukawa coupling constant and the deformation parameter. The effect of deformation on the membrane structure was revealed by the fermion massless modes precisely via Yukawa coupling.

The investigation of fermionic resonances is a subject possible here only numerically. However, from the qualitative character of the Shr\"odinger potentials we were able to observe that for large values of $f$ the behavior of localization of fermions is favored for larger values of $p$.
The presence of fermionic resonances was investigated for even parity wavefunctions both for left and right chiralities. We analyzed how the presence of the Yukawa coupling influenced our findings, with larger values of $f$ and $p$ favouring the increasing of resonances. Some resonances are extremely thin and more difficult to be found. The quantum mechanical supersymmetric character of the Shr\"odinger-like potentials for left and right chiral fermions guarantee that Dirac fermions are realized in the model and that the spectral lines for left and right fermions with same parity must alternate in the diagram. This is very helpful in the numerical process for finding resonances, serving as a guide for reducing the numerical step when a resonance line is not captured. The lightest modes in the spectrum correspond to the thinnest lines, showing that those modes couple strongly with the brane in comparison to the Kaluza-Klein modes with higher masses. This is expected as lightest modes have lower energy to escape from the brane Schr\"odinger potential. As the inverse of the peak width to half maximum is proportional to the lifetime of the resonance \cite{grs}, thinner peaks may correspond to particles in nature with sufficiently large lifetimes to be important in phenomenology.

The authors would like to thank CNPq and CAPES (Brazilian agencies) for financial support. Cruz and Almeida thank also FUNCAP; Gomes thanks also FAPEMA.

\end{document}